\documentclass{article}
\usepackage{amssymb}
\begin{document}

\def\Tr#1{{{\rm Tr}\,#1}}
\def\tr#1{{{\rm tr}\,#1}}
\def\Resp#1{{{\rm Res}^+_{#1}\,}}

\thispagestyle{empty}

\begin{flushright}
hep-th/0312262\\
AEI-2003-113 \\
\end{flushright}

\bigskip

\begin{center}

{\bf\Large Bulk Witten Indices from $D=10$ Yang Mills Integrals}

\bigskip\bigskip

{\bf T. Fischbacher\\}
\smallbreak
{\em Max-Planck-Institut f\"ur Gravitationsphysik,\\
  Albert-Einstein-Institut\\
  M\"uhlenberg 1, D-14476 Potsdam, Germany\\}
{\small {\tt tf@aei.mpg.de}}

\end{center}

\begin{abstract}
\noindent
Values for the bulk Witten indices for $D=10$ Yang-Mills integrals for
regular simple groups of rank 4 and 5 are calculated by employing the
BRST deformation technique by Moore, Nekrasov and Shatashvili.  The
results cannot be reconciled with the double assumption that the
number of normalizable ground states is given by certain simple
partition functions given by Kac and Smilga as well as that the
corresponding boundary term is always negative.
\end{abstract}

\section{Introduction}

Supersymmetric Yang-Mills theories in $D$ dimensions are described by an action of the form
\begin{equation}
S=\int d^Dx\,\left(-\frac{1}{4} F^{a\,\mu\nu} F^{a}_{\mu\nu}+\frac{i}{2}\bar\psi^{a}_\alpha\gamma^\mu_{\alpha\beta}\,D^{ab}_{\mu}\psi^{b}_\beta\right)
\end{equation}
where indices $a,b,\ldots$ correspond to the adjoint representation of
some semisimple compact Lie algebra, $\mu,\nu,\ldots$ are spacetime
vector indices (spatial indices will be denoted by $i,j,\ldots$), and
$\alpha,\beta,\ldots$ are spinorial spacetime indices; furthermore,
the nonabelian field strength is formed from the vector potential
$A^a_\mu$ via
\begin{equation}
F^{a}_{\mu\nu}=\partial_\mu A^{a}_\nu-\partial_\nu A^{a}_\mu+gf^a{}_{bc}A^b_\mu A^c_\nu,
\end{equation}
$f^a{}_{bc}$ being the gauge group structure constants,
and the gauge covariant derivative is given by
\begin{equation}
D^{ab}_{\mu}\psi^{b}_\beta=\delta^{ab}\partial_\mu\psi^{b}_\beta+gf^b{}_{cd}A^c_\mu\psi^d_\beta.
\end{equation}
The equality of the number of physical bosonic and fermionic degrees
of freedom required by supersymmetry furthermore forces the number of
transversal modes of the gauge field $A^a$, $D-2$, to be a power of
two, as is the size of the corresponding irreducible spinor
representation of the Lorentz algebra. One finds that it is in fact
possible here to implement supersymmetry in $D=3,4,6,10$. By
truncation to configurations that have no spatial dependency
(i.e. dimensional reduction to zero space dimensions, see
e.g.\cite{Danielsson:1996uw,Kabat:1996cu}), one obtains supersymmetric
quantum mechanics~\cite{Baake:1984ie,Claudson:1984th,Flume:1984mn,Smilga:jg}
with a Hamiltonian
\begin{equation}
H=\frac{1}{2}P_i^a P_i^a +\frac{g^2}4 f_{bca}f_{dea} A^b_i A^c_j A^d_i A^e_j +\frac{ig}{2}f_{abc}\psi^a_\alpha \Gamma^i_{\alpha\beta}\psi^b_\beta A_i^c
\end{equation}
with $P^a_i=-i\frac{\partial}{\partial A^a_i}$.

The $D=10$ case with gauge group $SU(N)$ regained a lot of popularity
after initial work by de Wit, Hoppe, and Nicolai~\cite{deWit:1988ig}
who showed the relevance of this particular model for the description
of the eleven-dimensional supermembrane (where $SU(N)$ appears as a
regularized version of the Lie group of area-preserving membrane
diffeomorphisms) through the so-called M(atrix) Theory
Conjecture~\cite{Banks:1996vh}, which states that this Hamiltonian of a
system of $N$ $D0$-branes of type IIA string theory should give a
complete, non-perturbative description of the dynamics of $M$-theory
in the light cone frame.\footnote{See, e.g. \cite{Taylor:2001vb} for
an overview}. Furthermore, these matrix models play an important role
in the IKKT model~\cite{Ishibashi:1996xs}, which may provide a
non-perturbative description of IIB superstring theory.

Gauge groups of other types are also of interest here, as realizations
of this model with $SO(N)$ and $Sp(2N)$ symmetry are given by systems
of $N$ type-IIA $D0$-branes moving in orientifold backgrounds,
cf.~\cite{Hanany:1999jy}.

One question of chief importance is that of normalizable zero energy
vacuum states of these models; this is difficult to settle as the
potentials have flat valleys that extend to infinity, and hence, the
corresponding ground states are at threshold. While it is exceedingly
difficult to try to explicitly solve the Schr\"odinger equation for
these models, it is already of great interest to know the {\em number}
of such normalizable ground states; for example, it is of crucial
importance to the Matrix Theory conjecture that there is exactly one
such state for every $N$ in the models derived from $SU(N)$ gauged
$\mathcal N=1$ SYM (which just corresponds to a bound state of $N$
$D0$-branes that appears as a graviton with $N$ units of momentum in
the compactified direction). While this is widely believed to be the
case by now, the situation is still much less clear for other gauge
groups.

As is nicely explained in~\cite{Sethi:2000zf}, the number of normalizable
ground states is given as the low temperature limit of the partition
function
\begin{equation}
\int dx\lim_{\beta\rightarrow\infty} \tr e^{-\beta H},
\end{equation}
but as the calculation of this quantity seems beyond reach for most
systems of interest, it appears more promising to try to calculate the
Witten index
\begin{equation}
I_w=n_B-n_F=\int dx\lim_{\beta\rightarrow\infty} \tr (-1)^F e^{-\beta H},
\end{equation}
instead (where $F$ is the fermion number), as this should also give
the number of normalizable ground states. In case of a discrete
spectrum, this partition function would be $\beta$-independent, so we
could as well take the limit $\beta\rightarrow 0$, which is accessible
in a perturbative calculation. For a continuous spectrum of the
Hamilton operator, this does not work in general, since supersymmetry
still pairs bosonic and fermionic modes, but the spectral density of
scattering states need not be equal. In this case, the `boundary' term
$I_1$ in the decomposition
\begin{equation}
\begin{array}{lcl}
I_w\kern-1em&=&\kern-1em I_0+I_1\\
&=&\kern-1em\displaystyle\lim_{R\rightarrow\infty}\lim_{\beta_0\rightarrow0}\int_{|x|<R}\!\!\!\!\!dx\left(\tr(-1)^Fe^{-\beta_0H(x,x)}+\int_{\beta_0}^\infty\!\!\!d\beta\,\partial_\beta\,\tr(-1)^Fe^{-\beta H}(x,x)\right)
\end{array}
\end{equation}
can acquire a nonzero value. While the technique of splitting the
integral into a `principal contribution' from the bulk term as well as
a `deficit contribution' which takes the form of a boundary
term~\cite{Sethi:2000zf} works remarkably well in many situations, as
one frequently finds that the boundary term is zero even if it {\em a
priori} does not seem to have to be
(e.g.~\cite{Sethi:1996kj,Alvarez-Gaume:1983at,Friedan:1983xr}), this
is not the case in the systems at hand. Nevertheless, the boundary
term has been calculated for $SU(2)$ in~\cite{Sethi:2000zf}, and
reasons (that are based on the (heuristic) assumption that for the
calculation of this deficit term the $D0$-branes can effectively be
treated as identical freely propagating particles, as in~\cite{Yi:1997eg})
have been given in~\cite{Green:1997tn} that this
deficit term should be
\begin{equation}
I_1^{D=10}(SU(N))=-\sum_{m|N,m>1}\frac{1}{m^2}
\end{equation}
for the $\mathcal N=1$ models derived from $D=10$ $SU(N)$ gauged SYM,
while one expects the value
\begin{equation}
I_1^{D=4}(SU(N))=I_1^{D=6}(SU(N))=-\frac{1}{N^2}
\end{equation}
for the $\mathcal N=4$ and $\mathcal N=2$ models derived from $D=4$,
resp. $D=6$ $SU(N)$ gauged SYM.

Employing the mass deformation method that has been developed 
in~\cite{Vafa:1994tf,Porrati:1997ej}, Kac and
Smilga~\cite{Kac:1999av} showed via group-theoretical means that under
the hypothesis that no large mass bound state becomes non-normalizable
as the zero-mass limit is taken, the number of normalizable ground
states should be given by certain simple partition functions:

\bigskip

\begin{equation}\label{KacSmilga}
\begin{array}{lcl}
I_w^{D=10}(SO(N))&=&\mbox{\# partitions of N into mutually distinct odd parts}\\
I_w^{D=10}(Sp(N))&=&\mbox{\# partitions of 2N into mutually distinct even parts}\\
I_w^{D=10}(G_2)&=&\phantom02\\
I_w^{D=10}(F_4)&=&\phantom04\\
I_w^{D=10}(E_6)&=&\phantom03\\
I_w^{D=10}(E_7)&=&\phantom06\\
I_w^{D=10}(E_8)&=&11\\
\end{array}
\end{equation}

Independent arguments that lead to the same $SO(N)$ and $Sp(2N)$
multiplicities have been presented in~\cite{Hanany:1999jy} which are based
on an analysis of the Hilbert space of a chiral fermion that is
constructed from D0 branes at an orientifold singularity.

The bulk term $I_0$ is -- at least in principle -- independently
accessible via a generalization~\cite{Staudacher:2000gx} of the BRST
deformation method which was devised by Moore, Nekrasov, and
Shatashvili~\cite{Moore:1998et} in order to greatly simplify the calculation
of the corresponding partition functions by adding terms to the action
which break all but one supersymmetry (so that the partition functions
do not change)\footnote{It was verified numerically 
in~\cite{Krauth:2000bv} for models obtained from $D=4$ SYM with low-rank
gauge groups that this method seemingly also works for other groups besides
$SU(N)$}. Concerning the boundary term, the method
of~\cite{Green:1997tn} has been generalized in~\cite{Kac:1999av} to other gauge
groups, but it was found that the expected Witten index $I_w=I_0+I_1$
could not be obtained that way~\cite{Staudacher:2000gx}, indicating a failure
of the assumption of the validity of the free Hamiltonian approach used 
in~\cite{Green:1997tn}. Nevertheless, for all the $D=4,6,10$ cases investigated
previously, the expected vacuum degeneracies support the
hypothesis~\cite{Staudacher:2000gx} that the boundary term is (with the
possible exception of $I_1^{D=10}(G_2)$) always a small number in the
interval $[-1;0]$, and hence a prediction of the Witten index should
be possible from the bulk index alone.

A first evaluation of $I_0^{D=10}$ for special orthogonal and
symplectic groups employing the Moore method has been performed
in~\cite{Staudacher:2000gx}; there as well as in further work on
$I_0^{D=4}$~\cite{Pestun:2002rr}, the algebraic bulkiness of bulk
Witten index calculations was pointed out, and indeed, even for the
$I_0^{D=4}$ case, going to ranks far beyond~$3$ already required
specialized term manipulation code to be written. In the following, we
want to review the operational issues of the application of the Moore
method to the even far more involved $D=10$ case and present
techniques (some conservative, some speculative) that allow the
calculation to be taken to regular simple groups of rank~4 and~5, and
also give arguments that such a direct approach is barely feasible for
simple groups of rank~$\ge 6$ employing currently available computer
technology.

\section{Computational aspects}

The heat kernel calculation~\cite{Yi:1997eg} of the bulk contribution
to $I_w$ requires evaluation of the partition function\footnote{Here and in the following,
we will use the conventions of~\cite{Staudacher:2000gx}}
\begin{equation}
\begin{array}{l}
\displaystyle I_0=\frac{1}{\mathcal F_G}\mathcal Z^{\mathcal N}_{D,G},\\
\displaystyle Z^{\mathcal N}_{D,G}=\int\prod_{a=1}^{{\rm dim}\,G}\prod_{\mu=1}^D\frac{dX_\mu^a}{(2\pi)^{1/2}}\prod_{\gamma=1}^{\mathcal N}d\psi_\gamma^a\times\\
\displaystyle \times\exp\left(\frac{1}{4g^2}\Tr[A_\mu,A_\nu][A_\mu,A_\nu]+\frac{1}{2g^2}\Tr\psi_\alpha[\Gamma^\mu_{\alpha\beta}X_\mu,\psi_\beta]\right)\\
\displaystyle =\int\prod_{a=1}^{{\rm dim}\,G}\prod_{\mu=1}^D\frac{dX_\mu^a}{(2\pi)^{1/2}}
 {\rm Pf}\,(-if^{bcd}\Gamma\cdot X^d)
\exp\left(\frac{1}{4g^2}\Tr[A_\mu,A_\nu][A_\mu,A_\nu]\right)\\

\end{array}
\end{equation}
where the fermionic degrees of freedom have been integrated out,
yielding a homogeneous Pfaffian, cf.~\cite{Krauth:1998xh}. These
partition functions $\mathcal Z^{\mathcal N}_{D,G}$, which come from
the reduction of SYM to zero dimensions (see
e.g.~\cite{Austing:2001ib}) have been dubbed `Yang-Mills
integrals'. The factor $\mathcal F_{G}$ that relates it to the bulk
Witten index is basically the effective gauge group volume,
see~\cite{Staudacher:2000gx} for details.

The BRST deformation technique of~\cite{Moore:1998et} greatly
simplifies this to the calculation of the integral~\cite{Staudacher:2000gx}

\begin{equation}
\begin{array}{l}
\displaystyle I_0^{D=10}(G)=\frac{|Z_G|}{|W_G|}\left(\frac{(E_1+E_2)(E_2+E_3)(E_3+E_1)}{E_1E_2E_3E_4}\right)^r\times\\
\times\displaystyle\oint\prod_{k=1}^r\frac{dx_k}{2\pi i}\frac{\tilde\Delta_G(0,\vec x)\tilde\Delta_G(E_1+E_2,\vec x)\tilde\Delta_G(E_2+E_3,\vec x)\tilde\Delta_G(E_3+E_1,\vec x)}{\tilde\Delta_G(E_1,\vec x)\tilde\Delta_G(E_2,\vec x)\tilde\Delta_G(E_3,\vec x)\tilde\Delta_G(E_4,\vec x)}\\
\mbox{with}\\
\displaystyle \tilde\Delta_G(E,\vec x)=\prod_{\vec\alpha\in\Delta_G}\left(\vec x\cdot\vec\alpha-E\right)
\end{array}
\end{equation}
where $r$ is the Lie group rank, $|Z_G|/|W_G|$ is the quotient of the
orders of the center and the Weyl group ($1/(2^rr!)$ for $B_r$,
$1/(2^{r-1}r!)$ for $C_r$, $1/(2^{r-2}r!)$ for $D_r$, 1/12 for $G_2$;
these are the only cases we are concerned with here), $E_{1,2,3,4}$
are auxiliary real quantities with $\sum_j E_j=0$ which the end result
will not depend on (at least for sufficiently generic values of $E_j$
where no `accidental' merging of poles/zeroes happens). $\Delta_G$ is
the set of roots of the Lie algebra $G$. Actually, as it stands these
integrals do not make much sense, as we have poles on the real axis of
integration. (Furthermore, they do not fall off fast enough towards
complex infinity to rigorously justify closing the integration
contour.) The correct (not yet fully justified,
see~\cite{Moore:1998et}) interpretation of these integrals rather is the
following algorithmic one:

\begin{enumerate}

\item Successively eliminate all $x_k$ `integration variables' in the term
\begin{equation}
\begin{array}{l}
\displaystyle\frac{|Z_G|}{|W_G|}\left(\frac{(E_1+E_2)(E_2+E_3)(E_3+E_1)}{E_1E_2E_3E_4}\right)^r\times\\
\times\displaystyle\lim_{\epsilon\rightarrow0+}\Resp{x_1}\ldots\Resp{x_r}\frac{\Delta_G(0,\vec x)\Delta_G(\tilde E_1+\tilde E_2,\vec x)}{\Delta_G(\tilde E_1,\vec x)\Delta_G(\tilde E_2,\vec x)}\times\\
\displaystyle\times\frac{\Delta_G(\tilde E_2+\tilde E_3,\vec x)\Delta_G(\tilde E_3+\tilde E_1,\vec x)}{\Delta_G(\tilde E_3,\vec x)\Delta_G(\tilde E_4,\vec x)}{\Biggr |}_{\tilde E_j=E_j+i\epsilon^j}
\end{array}
\end{equation}
where the $E_k, x_k$ are all treated as real, and $\Resp{x_j}$ picks up the residues with positive imaginary part only. (All the $x_k$ except the one being integrated out are treated as real.)

\item Substitute $E_4=-E_1-E_2-E_3$

\item Evaluate at generic values of $E_{1,2,3}$.
 (Alternatively: simplify to find that the result does not depend on $E_{1,2,3}$.)

\end{enumerate}

Due to the large number of factors in the denominator that occur in
these formulae for all but the smallest simple groups, one has to
resort to employing clever tricks to simplify the calculation, or
massive computer aid (or both).

While the use of symbolic manipulation programs like Maple or
Mathematica suggests itself for calculations like the ones at hand,
and indeed interesting results have been obtained that
way~\cite{Staudacher:2000gx}, the range to which these calculations can
be carried by employing such systems in a head-on approach is quite
limited, mostly due to the observation that (unlike the $D=4$ case) in
the evaluation of $(B,C,D)_n$ integrals, one seems to generically
encounter poles of order $(n-1)$ that require forming $(n-2)$th order
derivatives of very large products, leading to an explosion of the
number of terms generated. Nevertheless, it is obviously important to
try to obtain these values for as many gauge groups as possible in
order to check the validity of various assumptions that had to be made
in the calculation of Witten indices, and perhaps even make further
conjectures about analytical expressions, e.g. as
in~\cite{Pestun:2002rr} for $D=4$.

Even if one cannot do much about combinatoric explosion here, the
question nevertheless arises, whether -- considering the conceptual
simplicity of the problem -- one may be able to go a few steps further
by trying to make use of as much of the structure of the calculation
as possible in a dedicated program. While it is frequently possible to
outperform general-purpose symbolic algebra packages by three orders
of magnitude in term complexity with such an approach, more than one
rabbit has to be pulled out of the hat in order to achieve a
performance gain of one million or more, which we will see to be
necessary here. As the underlying techniques are of interest for a far
larger class of symbolic calculations, yet not widely known since
there is (to the author's knowledge) hardly any literature on the
relevant issues, we want to briefly present some of the fundamental
concepts one should be aware of when taking symbolic algebra to its
limits. Incidentally, the calculation at hand is an almost perfect
example to study these techniques.

Conventional symbolic algebra frequently (and especially in
calculations like these) wastes most of its time generating
intermediate quantities of very limited lifetime in dynamically
allocated memory which has to be reclaimed soon after. In the
following, we want to adhere to the convention to call this allocation
of dynamic memory which will not explicitly be reclaimed {\em
consing}\footnote{Named after the {\em cons} (pair), which is the most
ubiquitous building block of hierarchical data structures in LISP}. As
unnecessary consing forces frequent expensive calculations of
reachability graphs of in-memory objects (in order to identify
reclaimable space in the garbage collection process), it should as a
rule of thumb be avoided where possible. Furthermore, one should keep
in mind that for the presently dominant computer architecture the
processor's memory interface is an important bottleneck, and hence it
makes sense to try to find tight memory encodings for those pieces of
data on which most of the calculation operates so that cache stalls
are minimized.

One particularly striking feature of the calculation at hand is that
naively integrating out a single variable makes conventional term
manipulation programs first allocate huge numbers of almost similar
terms, which are then checked (in an overwhelming number of cases in
vain) for possible annihilations. This can be avoided by
re-structuring the calculation in such a way that instead of
generating and processing large amounts of individual mostly similar
(yet different enough that cancellations become quite rare) terms, one
and the same term backbone is destructively modified to consecutively
represent every single new term generated and then do further
processing on this backbone wherever possible. This technique is
particularly useful when it comes to processing higher-order
derivatives (where in addition we take care of performing the
iterations over places of factors where to derive in such a way that
re-occurring combinations of derivatives, as in
$(fg)''=f''g+f'g'+fg''+f'g'=f''g+fg''+2f'g'$ are only generated once).

One particular refinement for the present calculation is that one
generally can hardly avoid creating new terms by substitution, except
in the very last step (i.e. integrating out the last variable) where
this is indeed feasible. Since the last step is also executed most
frequently, this almost buys us an extra rank for free. (A slightly
nontrivial subtlety for the $D=10$ calculation here is that we have to
take care of the possibility of generating factors $E_1+E_2+E_3+E_4$
in both the numerator as well as the denominator. It may well happen
that while accumulating the factors of a term, one intermediately
encounters more such `powers of zero' in the denominator than in the
numerator.)

As calculations with exact numbers cause some systems to perform
excessive number consing, it is wise to try to avoid the exact
rational number data type for all those parts of a calculation where
one can use with impunity more limited data types that have fast
direct hardware support (like 32-bit integers)\footnote{Actually, most
systems providing a dynamic GC need to use a few tag bits to discern
between immediate and referenced (i.e. consed) values, so we typically
only can use 30 bit signed machine integers on 32-bit hardware}. For
our present calculation this means to represent the coefficients of
linear functions as machine integers. (Clearly, one has to take care
of proper handling of some `balancing denominators' that are generated
by this somewhat artificial treatment that requires to take least
common multiples at substitutions.) For all of the problems that are
reachable with present computer hardware, all of the substitutions of
such terms among themselves will not lead us outside the range of
these integer machine data types.

A further property of the calculations under study is that when
performed in the way described here, a large fraction of the
contributions are zero, and in many cases, this can be detected
somewhat easily without doing costly multiplications. Hence, it makes
sense to pre-scan a term for being zero before processing it wherever
applicable.

It is perhaps one of the less obvious properties of these Yang-Mills
integrals that one can do much better than to substitute particular
values for $E_{1\ldots 4}$ subject to the constraint $\sum_j E_j=0$
for the final evaluation; by bringing all factors to a proper
lexicographical normal form ($E_1$ contribution first if present, then
$E_2$ contribution, then $E_3$ contribution, all after having
substituted out $E_4$), one observes to obtain also the correct final
value by replacing every linear term by its leading
coefficient.\footnote{One of the more miraculous properties of these
integrals is that one also observes to obtain the correct final value
by replacing the sum in every linear factor by a product, i.e.
$\left(\sum_j c_j E_j\right)^p \rightarrow \left(\prod_j c_j
E_j\right)^p$; this was discovered through an intermediate programming
error.} While the largest calculations at present cannot be done without
employing this trick, we do a cross-check using the more conventional
method wherever this is possible. This is indicated in the last column
of our table of results.

As a special refinement, one notes that in the final gathering of
powers of linear factors, the most frequently encountered numerators
and denominators are $\pm1$ as well as small powers of 2. Indeed, a
further noticeable speedup can be achieved by treating these factors
special (remembering one overall power of 2 as well as the resulting
sign) in order to avoid unnecessary use of exact rational arithmetics.

As a basis for the implementation, the Objective Caml
system~\cite{Leroy} appears as very appealing, since it is highly
portable to a variety of different platforms, contains an optimizing
compiler that can generate compact standalone binaries, allows a very
smooth and easy two-way integration of C libraries and code; all these
qualities are highly desirable here especially since the problem at
hand suggests itself to massive parallelization, perhaps by making use
of donated computation time. (While rudimentary parallelization
support is present in our implementation and has proven its
usefulness, we do not implement such a large-scale scheme here, mainly
because it is expected that one would need a prohibitively large
number of volunteers to successfully do the next rank.) A further
advantage may be that Ocaml code generally is perceived as much less
alien by the uninitiated than LISP code. One further noteworthy issue
here is the quality of the built-in exact rational number arithmetics,
since there are all but obvious huge performance differences between
various implementations\footnote{For example, Gambit Scheme large
fraction arithmetics is roughly by a factor 2500 slower than the
excellent one present in CLISP, which even well outperforms e.g. that
of Maple.}. While the implementation present in Ocaml 3.07 is slower
than the one of CLISP by a factor of roughly 300, this is not a big
problem, since our techniques to reduce the use of fraction
arithmetics are powerful enough (at least when making use of the
evaluation shortcut described above) to make the amount of time spent
in arithmetics comparable to the time spent in other parts of the
program. CLISP would perhaps hardly be a viable alternative despite
its excellent arithmetics implementations, since it is only a
byte-code interpreter system.  As there are good reasons to make the
code with which these calculations have been performed publicly
available\footnote{first, as we are employing some algorithmically
nontrivial tricks, one cannot exclude the possibility of errors having
slipped in, hence it is provided for the sake of reproducibility and
checkability of all steps; second, there are many more
group-theoretical tricks one may want to exploit here and incorporate
into the existing code; third, one may want to carry the calculation
further as soon as more computer power becomes accessible, or by
employing massive parallelization; fourth, it may be instructive to
see the detailed implementation of some of the techniques described
above.}, it has been included in the {\tt arXiv.org} preprint upload
of this work\footnote{\tt http://www.arxiv.org/e-print/hep-th/0312262}.

\section{Results and discussion}

The following table shows our new as well as the previously known
values for the bulk contributions to the Witten index. Numerical
approximations as well as the expected values from the Kac/Smilga
hypothesis are also given. Calculation times refer to accumulated wall
clock times on a stepping-9 Intel Pentium 4 CPU, 2.4 GHz,
Hyperthreading enabled/ocaml 3.07, gcc 2.95, Linux kernel 2.4.24
student computer pool installation. (Values obtained on other
systems have been re-scaled appropriately.) Boldface indicates new values.

Some features that deserve special attention here are the particularly
large values for $SO(8)$ and $Sp(10)$ that would require $I_1<-1$, the
$SO(9)$ and $SO(11)$ values that would require $I_1>0$, and the
dramatic explosion of calculation time; beyond some `trivial' cases,
increasing the rank by one costs a factor (as a rule of thumb) of
roughly 1000 in CPU time in the interesting regime. Hence, it is
probably not yet feasible to try to attack rank~$6$ for the
$(B,C,D)_n$ groups. As the $A_n$ bulk Indices are just the expected
ones, and as the $(B,C,D)_n$ are of the right magnitude, with
denominators being powers of two (this is not the case at all for
individual summands), there is good reason to believe in both the
validity of this approach and its implementation. However, the
calculation of the $C_4$ index by using explicit values for
$E_{1,2,3}$ seems to fail systematically for yet unknown reasons,
despite a very careful analysis of the code. (This may be related to
yet another bug in the ocaml compiler; this assumption is also fueled
by the observation that part of the $D_6$ calculation causes memory
violations that should not be possible at all in pure ocaml for a very
specific combination of processors and optimization flags.)

\bigbreak

{\small
\begin{tabular}{|lccclc|}
\hline
Group&$I_0^{D=10}$&num.&expected&approx. calc.&cross-check\\
&&approx.&$I_w$&time (sec)&\\
\hline
&&&&&\\
$SO(\phantom03)\;(B_1)$&$\frac{5}{4}$&$1.2500$&$1$&$2.6\cdot10^{-4}$&{\rm ok}\\&&&&&\\
$SO(\phantom05)\;(B_2)$&$\frac{81}{64}$&$1.2656$&$1$&$1.1\cdot10^{-2}$&{\rm ok}\\&&&&&\\
$SO(\phantom07)\;(B_3)$&$\frac{325}{256}$&$1.26953$&$1$&$7.8\cdot10^{-1}$&{\rm ok}\\&&&&&\\
$SO(\phantom09)\;(B_4)$&${\bf \frac{15925}{8192}}$&$1.8671$&$2$&$6.6\cdot10^{2}$&{\rm ok}\\&&&&&\\
$SO(11)\;(B_5)$&${\bf \frac{59049}{32768}}$&$1.8020$&$2$&$2.8\cdot10^{6}$&\\&&&&&\\
\hline
&&&&&\\
$SO(\phantom04)\;(D_2)$&$\frac{25}{16}$&$1.5625$&$1$&$3.4\cdot10^{-3}$&{\rm ok}\\&&&&&\\
$SO(\phantom06)\;(D_3)$&$\frac{21}{16}$&$1.3125$&$1$&$1.9\cdot10^{-1}$&{\rm ok}\\&&&&&\\
$SO(\phantom08)\;(D_4)$&${\bf \frac{6885}{2048}}$&$3.3618$&$2$&$1.5\cdot10^{2}$&{\rm ok}\\&&&&&\\
$SO(10)\;(D_5)$&${\bf \frac{3025}{1024}}$&$2.9541$&$2$&$2.1\cdot10^{5}$&\\&&&&&\\
\hline
&&&&&\\
$Sp(\phantom02)\;(C_1)$&$\frac{5}{4}$&$1.2500$&$1$&$3.2\cdot10^{-4}$&{\rm ok}\\&&&&&\\
$Sp(\phantom04)\;(C_2)$&$\frac{81}{64}$&$1.2656$&$1$&$9.6\cdot10^{-3}$&{\rm ok}\\&&&&&\\
$Sp(\phantom06)\;(C_3)$&$\frac{1175}{512}$&$2.2949$&$2$&$7.7\cdot10^{-1}$&{\rm ok}\\&&&&&\\
$Sp(\phantom08)\;(C_4)$&${\bf \frac{42667}{16384}}$&$2.6042$&$2$&$5.2\cdot10^{2}$&{\rm failed}\\&&&&&\\
$Sp(10)\;(C_5)$&${\bf \frac{583755}{131072}}$&$4.4537$&$3$&$1.9\cdot10^{6}$&\\
&&&&&\\
\hline
&&&&&\\
$SU(\phantom02)\;(A_1)$&$\frac{5}{4}$&$1.2500$&$1$&$3.4\cdot10^{-4}$&{\rm ok}\\&&&&&\\
$SU(\phantom03)\;(A_2)$&$\frac{10}{9}$&$1.1111$&$1$&$8.5\cdot10^{-3}$&{\rm ok}\\&&&&&\\
$SU(\phantom04)\;(A_3)$&$\frac{21}{16}$&$1.3125$&$1$&$3.7\cdot10^{-1}$&{\rm ok}\\&&&&&\\
$SU(\phantom05)\;(A_4)$&$\frac{26}{25}$&$1.0400$&$1$&$4.7\cdot10^{1}$&{\rm ok}\\&&&&&\\
$SU(\phantom06)\;(A_5)$&$\frac{25}{18}$&$1.3889$&$1$&$1.4\cdot10^{4}$&\\&&&&&\\
$SU(\phantom07)\;(A_6)$&$\frac{50}{49}$&$1.0204$&$1$&$1.1\cdot10^{7}$&\\&&&&&\\
\hline
&&&&&\\
$G_2$&$\frac{1375}{864}$&$1.5914$&$2$&$4.5\cdot10^{-2}$&{\rm ok}\\&&&&&\\
\hline
\end{tabular}
}

\bigbreak
\bigbreak

All in all, the new data clearly show that for other gauge groups than
$SU(N)$, the issue of the number of vacuum states is not well
understood, and more work has to be put into the determination of
Witten indices. In particular, we do not have at present a useful
theory to calculate the boundary contributions, and it is well
conceivable that some of the assumptions behind $(\ref{KacSmilga})$
may be violated.

{
\paragraph{Acknowledgments}\hfill\bigbreak

\noindent I want to thank Matthias Staudacher for introducing me to this
interesting problem, and also for many helpful comments and
discussions, the administrative staff of {\tt
cip.physik.uni-munchen.de}, where some of the calculations have been
performed, as well as Steffen Grunewald for letting me submit some
calculation jobs to the Geo600 Merlin cluster at AEI. Furthermore, I
want to thank the Debian ocaml package maintainer, Sven Luther, for
his for his prompt handling of bug reports, and Damien Doligez for
fixing an ocaml compiler bug~\cite{ocamlbug} that was discovered
through this work.}

\end{document}